\documentclass{acm_proc_article-sp}
\usepackage[utf8]{inputenc}
\usepackage[T1]{fontenc}
\usepackage{fixltx2e}
\usepackage{graphicx}
\usepackage{longtable}
\usepackage{float}
\usepackage{wrapfig}
\usepackage{rotating}
\usepackage[normalem]{ulem}
\usepackage{amsmath}
\usepackage{textcomp}
\usepackage{marvosym}
\usepackage{wasysym}
\usepackage{amssymb}
\usepackage{hyperref}
\DeclareTextFontCommand{\texttt}{\ttfamily\hyphenchar\font=45\relax}

\author{Christophe Rhodes, Jan Moringen, David Lichteblau}
\date{\today}
\title{Generalizers: New Metaobjects for Generalized Dispatch}
\hypersetup{
  pdfkeywords={},
  pdfsubject={},
  pdfcreator={Emacs 24.3.1 (Org mode 8.2.4)}}
\begin{document}

\numberofauthors{3}
\author{
\alignauthor Christophe Rhodes\\
  \affaddr{Department of Computing}\\
  \affaddr{Goldsmiths, University of London}\\
  \affaddr{London SE14 6NW}\\
  \email{c.rhodes@gold.ac.uk}
\alignauthor Jan Moringen\\
  \affaddr{Universität Bielefeld}\\
  \affaddr{Technische Fakultät}\\
  \affaddr{33594 Bielefeld}\\
  \email{jmoringe@techfak.uni-bielefeld.de}
\alignauthor David Lichteblau\\
  \affaddr{ZenRobotics Ltd}\\
  \affaddr{Vilhonkatu 5 A}\\
  \affaddr{FI-00100 Helsinki}\\
  \email{david@lichteblau.com}
}
\maketitle
\begin{abstract}
This paper introduces a new metaobject, the generalizer, which
complements the existing specializer metaobject.  With the help of
examples, we show that this metaobject allows for the efficient
implementation of complex non-class-based dispatch within the
framework of existing metaobject protocols.  We present our
modifications to the generic function invocation protocol from the
\emph{Art of the Metaobject Protocol}; in combination with previous work,
this produces a fully-functional extension of the existing mechanism
for method selection and combination, including support for method
combination completely independent from method selection.  We discuss
our implementation, within the SBCL implementation of Common Lisp, and
in that context compare the performance of the new protocol with the
standard one, demonstrating that the new protocol can be tolerably
efficient.
\end{abstract}

\begin{flushleft}
Report-No.:~\url{http://eprints.gold.ac.uk/id/eprint/9924}
\end{flushleft}

\section{Introduction}
\label{sec-1}
The revisions to the original Common Lisp language \cite{CLtL}
included the detailed specification of an object system, known as
the Common Lisp Object System (CLOS), which was eventually
standardized as part of the ANSI Common Lisp standard \cite{CLtS}.
The object system as presented to the standardization committee was
formed of three chapters.  The first two chapters covered programmer
interface concepts and the functions in the programmer interface
\cite[Chapter 28]{CLtL2} and were largely incorporated into the
final standard; the third chapter, covering a Metaobject Protocol
(MOP) for CLOS, was not.

Nevertheless, the CLOS MOP continued to be developed, and the
version documented in \cite{AMOP} has proven to be a reasonably
robust design.  While many implementations have derived their
implementations of CLOS from either the Closette illustrative
implementation in \cite{AMOP}, or the Portable Common Loops
implementation of CLOS from Xerox Parc, there have been largely
from-scratch reimplementations of CLOS (in CLISP\footnote{GNU CLISP, at \url{http://www.clisp.org/}} and
CCL\footnote{Clozure Common Lisp, at \url{http://ccl.clozure.com/}}, at least) incorporating substantial fractions of the
Metaobject Protocol as described.

\begin{figure}[htb]
\centering
\includegraphics[width=.9\linewidth]{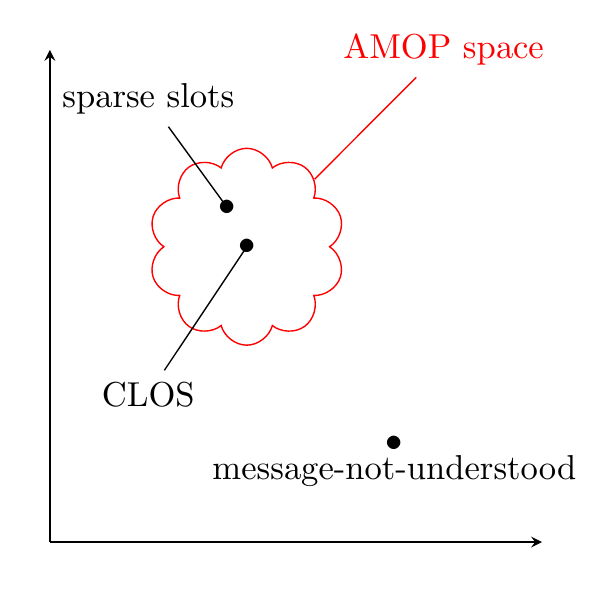}
\caption{\label{fig:mopdesign}MOP Design Space}
\end{figure}

Although it has stood the test of time, the CLOS MOP is neither
without issues (e.g. semantic problems with \texttt{make-method-lambda}
\cite{Costanza.Herzeel:2008}; useful functions such as
\texttt{compute-effective-slot-definition-initargs} being missing from the
standard) nor is it a complete framework for the metaprogrammer to
implement all conceivable variations of object-oriented behaviour.
While metaprogramming offers some possibilities for customization of
the object system behaviour, those possibilities cannot extend
arbitrarily in all directions (conceptually, if a given object
system is a point in design space, then a MOP for that object system
allows exploration of a region of design space around that point;
see figure \ref{fig:mopdesign}).  In the case of the CLOS MOP, there is
still an expectation that functionality is implemented with methods
on generic functions, acting on objects with slots; it is not
possible, for example, to transparently implement support for
“message not understood” as in the message-passing paradigm, because
the analogue of messages (generic functions) need to be defined
before they are used.

Nevertheless, the MOP is flexible, and is used for a number of
things, including: documentation generation (where introspection in
the MOP is used to extract information from a running system\footnote{as in many of the systems surveyed at
\url{https://sites.google.com/site/sabraonthehill/lisp-document-generation-apps}});
object-relational mapping\footnote{e.g. CLSQL, at \url{http://clsql.b9.com/}} and other approaches to object
persistence \cite{Paepke:1988}; alternative backing stores for slots
(hash-tables \cite{Kiczales.etal:1993} or symbols
\cite{Costanza.Hirschfeld:2005}); and programmatic construction of
metaobjects, for example for interoperability with other language
runtimes' object systems.

One area of functionality where there is scope for customization by
the metaprogrammer is in the mechanics and semantics of method
applicability and dispatch.  While in principle AMOP allows
customization of dispatch in various different ways (the
metaprogrammer can define methods on protocol functions such as
\texttt{compute-applicable-methods},
\texttt{compute-applicable-methods-using-classes}), for example, in
practice implementation support for this was weak until relatively
recently\footnote{the \textsf{Closer to MOP} project, at
\url{http://common-lisp.net/project/closer/}, attempts to harmonize the
different implementations of the metaobject protocol in Common
Lisp.}.

Another potential mechanism for customizing dispatch is implicit in
the class structure defined by AMOP: standard specializer objects
(instances of \texttt{class} and \texttt{eql-specializer}) are generalized
instances of the \texttt{specializer} protocol class, and in principle
there are no restrictions on the metaprogrammer constructing
additional subclasses.  Previous work \cite{Newton.Rhodes:2008} has
explored the potential for customizing generic function dispatch
using extended specializers, but there the metaprogrammer must
override the entirety of the generic function invocation protocol
(from \texttt{compute-discriminating-function} on down), leading to toy
implementations and duplicated effort.

\begin{figure}[htb]
\centering
\includegraphics[width=.9\linewidth]{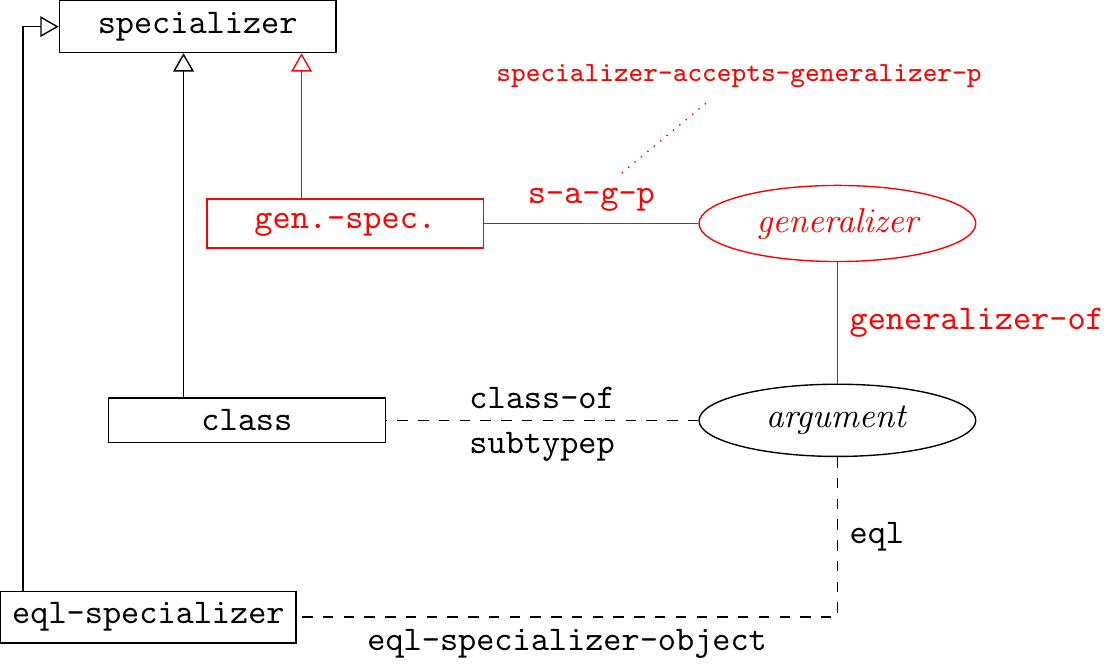}
\caption{\label{fig:dispatch}Dispatch Comparison}
\end{figure}

This paper introduces a protocol for efficient and controlled
handling of new subclasses of \texttt{specializer}.  In particular, it
introduces the \texttt{generalizer} protocol class, which generalizes the
return value of \texttt{class-of} in method applicability computation, and
allows the metaprogrammer to hook into cacheing schemes to avoid
needless recomputation of effective methods for sufficiently similar
generic function arguments (See Figure~\ref{fig:dispatch}).

The remaining sections in this paper can be read in any order.  We
give some motivating examples in section \ref{sec-2}, including
reimplementations of examples from previous work, as well as
examples which are poorly supported by previous protocols.  We
describe the protocol itself in section \ref{sec-3}, describing each
protocol function in detail and, where applicable, relating it to
existing protocol functions within the CLOS MOP.  We survey related
work in more detail in section \ref{sec-4}, touching on work on
customized dispatch schemes in other environments.  Finally, we draw
our conclusions from this work, and indicate directions for further
development, in section \ref{sec-5}; reading that section before the
others indicates substantial trust in the authors' work.

\section{Examples}
\label{sec-2}
In this section, we present a number of examples of dispatch
implemented using our protocol, which we describe in section
\ref{sec-3}.  For reasons of space, the metaprogram code examples in
this section do not include some of the necessary support code to
run; complete implementations of each of these cases, along with the
integration of this protocol into the SBCL implementation
\cite{Rhodes:2008} of Common Lisp, are included in the authors'
repository\footnote{the tag \texttt{els2014-submission} in
\url{http://christophe.rhodes.io/git/specializable.git} corresponds to the
code repository at the point of submitting this paper.}.

A note on terminology: we will attempt to distinguish between the
user of an individual case of generalized dispatch (the
“programmer”), the implementor of a particular case of generalized
dispatch (the “metaprogrammer”), and the authors as the designers
and implementors of our generalized dispatch protocol (the
“metametaprogrammer”, or more likely “we”).
\subsection{CONS specializers}
\label{sec-2-1}
One motivation for the use of generalized dispatch is in an
extensible code walker: a new special form can be handled simply by
writing an additional method on the walking generic function,
seamlessly interoperating with all existing methods. In this
use-case, dispatch is performed on the first element of lists.
Semantically, we allow the programmer to specialize any argument of
methods with a new kind of specializer, \texttt{cons-specializer}, which
is applicable if and only if the corresponding object is a \texttt{cons}
whose \texttt{car} is \texttt{eql} to the symbol associated with the
\texttt{cons-specializer}; these specializers are more specific than the
\texttt{cons} class, but less specific than an \texttt{eql-specializer} on any
given \texttt{cons}.

The programmer code using these specializers is unchanged from
\cite{Newton.Rhodes:2008}; the benefits of the protocol described
here are: that the separation of concerns is complete – method
selection is independent of method combination – and that the
protocol allows for efficient implementation where possible, even
when method selection is customized.  In an application such as
walking source code, we would expect to encounter special forms
(distinguished by particular atoms in the \texttt{car} position) multiple
times, and hence to dispatch to the same effective method
repeatedly.  We discuss the efficiency aspects of the protocol in
more detail in section \ref{sec-3-1-2}; we present the metaprogrammer
code to implement the \texttt{cons-specializer} below.

\begin{verbatim}
(defclass cons-specializer (specializer)
  ((%car :reader %car :initarg :car)))
(defclass cons-generalizer (generalizer)
  ((%car :reader %car :initarg :car)))
(defmethod generalizer-of-using-class
    ((gf cons-generic-function) arg)
  (typecase arg
    ((cons symbol)
     (make-instance 'cons-generalizer
                    :car (car arg)))
    (t (call-next-method))))
(defmethod generalizer-equal-hash-key
    ((gf cons-generic-function)
     (g cons-generalizer))
  (%car g))
(defmethod specializer-accepts-generalizer-p
    ((gf cons-generic-function)
     (s cons-specializer)
     (g cons-generalizer))
  (if (eql (%car s) (%car g))
      (values t t)
      (values nil t)))
(defmethod specializer-accepts-p
    ((s cons-specializer) o)
  (and (consp o) (eql (car o) (%car s))))
\end{verbatim}

The code above shows a minimal use of our protocol.  We have elided
some support code for parsing and unparsing specializers, and for
handling introspective functions such as finding generic functions for
a given specializer.  We have also elided methods on the protocol
functions \texttt{specializer<} and \texttt{same-specializer-p}; for
\texttt{cons-specializer} objects, specializer ordering is trivial, as only
one \texttt{cons-specializer} (up to equality) can ever be applicable to any
given argument.  See section \ref{sec-2-3} for a case where specializer
ordering is non-trivial.

As in \cite{Newton.Rhodes:2008}, the programmer can use these
specializers to implement a modular code walker, where they define one
method per special operator.  We show two of those methods below, in
the context of a walker which checks for unused bindings and uses of
unbound variables.

\begin{verbatim}
(defgeneric walk (form env stack)
  (:generic-function-class cons-generic-function))
(defmethod walk
    ((expr (cons lambda)) env call-stack)
  (let ((lambda-list (cadr expr))
        (body (cddr expr)))
    (with-checked-bindings
        ((bindings-from-ll lambda-list)
         env call-stack)
      (dolist (form body)
        (walk form env (cons form call-stack))))))
(defmethod walk
    ((expr (cons let)) env call-stack)
  (flet ((let-binding (x)
           (walk (cadr x) env
                 (cons (cadr x) call-stack))
           (cons (car x)
                 (make-instance 'binding))))
    (with-checked-bindings
        ((mapcar #'let-binding (cadr expr))
          env call-stack)
      (dolist (form (cddr expr))
        (walk form env (cons form call-stack))))))
\end{verbatim}

Note that in this example there is no strict need for
\texttt{cons-specializer} and \texttt{cons-generalizer} to be distinct classes.
In standard generic function dispatch, the \texttt{class} functions both
as the specializer for methods and as the generalizer for generic
function arguments; we can think of the dispatch implemented by
\texttt{cons-specializer} objects as providing for subclasses of the
\texttt{cons} class distinguished by the \texttt{car} of the \texttt{cons}.  This
analogy also characterizes those use cases where the metaprogrammer
could straightforwardly use filtered dispatch
\cite{Costanza.etal:2008} to implement their dispatch semantics.
We will see in section \ref{sec-2-3} an example of a case where filtered
dispatch is incapable of straightforwardly expressing the dispatch,
but first we present our implementation of the motivating case from
\cite{Costanza.etal:2008}.

\subsection{SIGNUM specializers}
\label{sec-2-2}
Our second example of the implementation and use of generalized
specializers is a reimplementation of one of the examples in
\cite{Costanza.etal:2008}: specifically, the factorial function.
Here, dispatch will be performed based on the \texttt{signum} of the
argument, and again, at most one method with a \texttt{signum} specializer
will be applicable to any given argument, which makes the structure
of the specializer implementation very similar to the \texttt{cons}
specializers in the previous section.

The metaprogrammer has chosen in the example below to compare
signum values using \texttt{=}, which means that a method with
specializer \texttt{(signum 1)} will be applicable to positive
floating-point arguments (see the first method on
\texttt{specializer-accepts-generalizer-p} and the method on
\texttt{specializer-accepts-p} below).  This leads to one subtle
difference in behaviour compared to that of the \texttt{cons}
specializers: in the case of \texttt{signum} specializers, the \emph{next}
method after any \texttt{signum} specializer can be different, depending
on the class of the argument.  This aspect of the dispatch is
handled by the second method on \texttt{specializer-accepts-generalizer-p}
below.

\begin{verbatim}
(defclass signum-specializer (specializer)
  ((%signum :reader %signum :initarg :signum)))
(defclass signum-generalizer (generalizer)
  ((%signum :reader %signum :initarg :signum)))
(defmethod generalizer-of-using-class
    ((gf signum-generic-function) (arg real))
  (make-instance 'signum-generalizer
                 :signum (signum arg)))
(defmethod generalizer-equal-hash-key
    ((gf signum-generic-function)
     (g signum-generalizer))
  (%signum g))
(defmethod specializer-accepts-generalizer-p
    ((gf signum-generic-function)
     (s signum-specializer)
     (g signum-generalizer))
  (if (= (%signum s) (%signum g))
      (values t t)
      (values nil t)))

(defmethod specializer-accepts-generalizer-p
    ((gf signum-generic-function)
     (s specializer)
     (g signum-generalizer))
  (specializer-accepts-generalizer-p
   gf s (class-of (%signum g))))

(defmethod specializer-accepts-p
    ((s signum-specializer) o)
  (and (realp o) (= (%signum s) (signum o))))
\end{verbatim}

Given these definitions, and once again some more straightforward
ones elided for reasons of space, the programmer can implement the
factorial function as follows:

\begin{verbatim}
(defgeneric fact (n)
  (:generic-function-class signum-generic-function))
(defmethod fact ((n (signum 0))) 1)
(defmethod fact ((n (signum 1))) (* n (fact (1- n))))
\end{verbatim}

The programmer does not need to include a method on \texttt{(signum -1)},
as the standard \texttt{no-applicable-method} protocol will automatically
apply to negative real or non-real arguments.

\subsection{Accept HTTP header specializers}
\label{sec-2-3}
In this section, we implement a non-trivial form of dispatch.  The
application in question is a web server, and specifically to allow
the programmer to support RFC 2616 \cite{rfc2616} content
negotiation, of particular interest to publishers and consumers of
REST-style Web APIs.

The basic mechanism in content negotiation is as follows: the web
client sends an HTTP request with an \texttt{Accept} header, which is a
string describing the media types it is willing to receive as a
response to the request, along with numerical preferences.  The web
server compares these stated client preferences with the resources
it has available to satisfy this request, and sends the best
matching resource in its response.

For example, a graphical web browser might send an \texttt{Accept} header
of \texttt{text/html,application/xml;q=0.9,*/*;q=0.8} for a request of a
resource typed in to the URL bar.  This should be interpreted as
meaning that: if the server can provide content of type \texttt{text/html}
(i.e. HTML) for that resource, then it should do so.  Otherwise, if
it can provide \texttt{application/xml} content (i.e. XML of any schema),
then that should be provided; failing that, any other content type
is acceptable.

In the case where there are static files on the filesystem, and the
web server must merely select between them, there is not much more
to say.  However, it is not unusual for a web service to be backed
by some other form of data, and responses computed and sent on the
fly, and in these circumstances the web server must compute which
of its known output formats it can use to satisfy the request
before actually generating the best matching response.  This can be
modelled as one generic function responsible for generating the
response, with methods corresponding to content-types -- and the
generic function must then perform method selection against the
request's \texttt{Accept} header to compute the appropriate response.

The \texttt{accept-specializer} below implements this dispatch.  It depends
on a lazily-computed \texttt{tree} slot to represent the information in
the accept header (generated by \texttt{parse-accept-string}), and a
function \texttt{q} to compute the (defaulted) preference level for a
given content-type and \texttt{tree}; then, method selection and ordering
involves finding the \texttt{q} for each \texttt{accept-specializer}'s content
type given the \texttt{tree}, and sorting them according to the preference
level.

\begin{verbatim}
(defclass accept-specializer (specializer)
  ((media-type :initarg :media-type :reader media-type)))
(defclass accept-generalizer (generalizer)
  ((header :initarg :header :reader header)
   (tree)
   (next :initarg :next :reader next)))
(defmethod generalizer-equal-hash-key
    ((gf accept-generic-function)
     (g accept-generalizer))
  `(accept-generalizer ,(header g)))
(defmethod specializer-accepts-generalizer-p
    ((gf accept-generic-function)
     (s accept-specializer)
     (g accept-generalizer))
  (values (q (media-type s) (tree g)) t))
(defmethod specializer-accepts-generalizer-p
    ((gf accept-generic-function)
     (s specializer)
     (g accept-generalizer))
  (specializer-accepts-generalizer-p
   gf s (next g)))

(defmethod specializer<
    ((gf accept-generic-function)
     (s1 accept-specializer)
     (s2 accept-specializer)
     (g accept-generalizer))
  (let ((m1 (media-type s1))
        (m2 (media-type s2))
        (tree (tree g)))
    (cond
      ((string= m1 m2) '=)
      (t (let ((q1 (q m1 tree)))
               (q2 (q m2 tree))))
           (cond
             ((= q1 q2) '=)
             ((< q1 q2) '>)
             (t '<))))))
\end{verbatim}

The metaprogrammer can then add support for objects representing
client requests, such as instances of the \texttt{request} class in the
Hunchentoot\footnote{Hunchentoot is a web server written in Common Lisp, allowing
the user to write handler functions to compute responses to requests;
\url{http://weitz.de/hunchentoot/}} web server, by translating these into
\texttt{accept-generalizer} instances.  The code below implements this, by
defining the computation of a \texttt{generalizer} object for a given
request, and specifying how to compute whether the specializer
accepts the given request object (\texttt{q} returns a number between 0
and 1 if any pattern in the \texttt{tree} matches the media type, and
\texttt{nil} if the media type cannot be matched at all).

\begin{verbatim}
(defmethod generalizer-of-using-class
    ((gf accept-generic-function)
     (arg tbnl:request))
  (make-instance 'accept-generalizer
                 :header (tbnl:header-in :accept arg)
                 :next (call-next-method)))
(defmethod specializer-accepts-p
    ((s accept-specializer)
     (o tbnl:request))
  (let* ((accept (tbnl:header-in :accept o))
         (tree (parse-accept-string accept))
         (q (q (media-type s) tree)))
    (and q (> q 0))))
\end{verbatim}

This dispatch cannot be implemented using filtered dispatch, except
by generating anonymous classes with all the right mime-types as
direct superclasses in dispatch order; the filter would generate
\begin{verbatim}
(ensure-class nil :direct-superclasses
 '(text/html image/webp ...))
\end{verbatim}
and dispatch would operate using those anonymous classes.  While
this is possible to do, it is awkward to express content-type
negotiation in this way, as it means that the dispatcher must know
about the universe of mime-types that clients might declare that
they accept, rather than merely the set of mime-types that a
particular generic function is capable of serving; handling
wildcards in accept strings is particularly awkward in the
filtering paradigm.

Note that in this example, the method on \texttt{specializer<} involves a
non-trivial ordering of methods based on the \texttt{q} values specified
in the accept header (whereas in sections \ref{sec-2-1} and \ref{sec-2-2} only a
single extended specializer could be applicable to any given
argument).

Also note that the accept specializer protocol is straightforwardly
extensible to other suitable objects; for example, one simple
debugging aid is to define that an \texttt{accept-specializer} should be
applicable to \texttt{string} objects.  This can be done in a modular
fashion (see the code below, which can be completely disconnected
from the code for Hunchentoot request objects), and generalizes to
dealing with multiple web server libraries, so that
content-negotiation methods are applicable to each web server's
request objects.

\begin{verbatim}
(defmethod generalizer-of-using-class
    ((gf accept-generic-function)
     (s string))
  (make-instance 'accept-generalizer
                 :header s
                 :next (call-next-method)))
(defmethod specializer-accepts-p
    ((s accept-specializer) (o string))
  (let* ((tree (parse-accept-string o))
         (q (q (media-type s) tree)))
    (and q (> q 0))))
\end{verbatim}

The \texttt{next} slot in the \texttt{accept-generalizer} is used to deal with
the case of methods specialized on the classes of objects as well
as on the acceptable media types; there is a method on
\texttt{specializer-accepts-generalizer-p} for specializers that are not
of type \texttt{accept-specializer} which calls the generic function again
with the next generalizer, so that methods specialized on the
classes \texttt{tbnl:request} and \texttt{string} are treated as applicable to
corresponding objects, though less specific than methods with
\texttt{accept-specializer} specializations.

\section{Protocol}
\label{sec-3}

In section \ref{sec-2}, we have seen a number of code fragments as
partial implementations of particular non-standard method dispatch
strategies, using \texttt{generalizer} metaobjects to mediate between the
methods of the generic function and the actual arguments passed to
it.  In section \ref{sec-3-1}, we go into more detail
regarding these \texttt{generalizer} metaobjects, describing the generic
function invocation protocol in full, and showing how this protocol
allows a similar form of effective method cacheing as the standard
one does.  In section \ref{sec-3-2}, we show the results
of some simple performance measurements on our implementation of
this protocol in the SBCL implementation \cite{Rhodes:2008} of
Common Lisp to highlight the improvement that this protocol can
bring over a naïve implementation of generalized dispatch, as well
as to make the potential for further improvement clear.

\subsection{Generalizer metaobjects}
\label{sec-3-1}

\subsubsection{Generic function invocation}
\label{sec-3-1-1}
As in the standard generic function invocation protocol, the
generic function's actual functionality is provided by a
discriminating function.  The functionality described in this
protocol is implemented by having a distinct subclass of
\texttt{standard-generic-function}, and a method on
\texttt{compute-discriminating-function} which produces a custom
discriminating function.  The basic outline of the discriminating
function is the same as the standard one: it must first compute the
set of applicable methods given particular arguments; from that, it
must compute the effective method by combining the methods
appropriately according to the generic function's method
combination; finally, it must call the effective method with the
arguments.

Computing the set of applicable methods is done using a pair of
functions: \texttt{compute-applicable-methods}, the standard metaobject
function, and a new function
\texttt{compute-applicable-methods-using-generalizers}.  We define a
custom method on \texttt{compute-applicable-methods} which tests the
applicability of a particular specializer against a given argument
using \texttt{specializer-accepts-p}, a new protocol function with
default implementations on \texttt{class} and \texttt{eql-specializer} to
implement the expected behaviour.  To order the methods, as
required by the protocol, we define a pairwise comparison operator
\texttt{specializer<} which defines an ordering between specializers for
a given generalizer argument (remembering that even in standard
CLOS the ordering between \texttt{class} specializers can change
depending on the actual class of the argument).

The new \texttt{compute-applicable-methods-using-generalizers} is the
analogue of the MOP's \texttt{compute-applicable-methods-using-classes}.
Instead of calling it with the \texttt{class-of} each argument, we
compute the generalizers of each argument using the new function
\texttt{generalizer-of-using-class} (where the \texttt{-using-class} refers to
the class of the generic function rather than the class of the
object), and call \texttt{compute-applicable-methods-using-generalizers}
with the generic function and list of generalizers.  As with
\texttt{compute-applicable-methods-using-classes}, a secondary return
value indicates whether the result of the function is definitive
for that list of generalizers.

Thus, in generic function invocation, we first compute the
generalizers of the arguments; we compute the ordered set of
applicable methods, either from the generalizers or (if that is
not definitive) from the arguments themselves; then the normal
effective method computation and call can occur.  Unfortunately,
the nature of an effective method function is not specified, so we
have to reach into implementation internals a little in order to
call it, but otherwise the remainder of the generic function
invocation protocol is unchanged from the standard one.  In
particular, method combination is completely unchanged;
programmers can choose arbitrary method combinations, including
user-defined long form combinations, for their generic functions
involving generalized dispatch.
\subsubsection{Effective method memoization}
\label{sec-3-1-2}
The potential efficiency benefit to having \texttt{generalizer}
metaobjects lies in the use of
\texttt{compute-applicable-methods-using-generalizers}.  If a particular
generalized specializer accepts a variety of objects (such as the
\texttt{signum} specializer accepting all reals with a given sign, or the
\texttt{accept} specializer accepting all HTTP requests with a particular
\texttt{Accept} header), then there is the possibility of cacheing and
reusing the results of the applicable and effective method
computation.  If the computation of the applicable method from
\texttt{compute-applicable-methods-using-generalizers} is definitive,
then the ordered set of applicable methods and the effective
method can be cached.

One issue is what to use as the key for that cache.  We cannot use
the generalizers themselves, as two generalizers that should be
considered equal for cache lookup will not compare as \texttt{equal} –
and indeed even the standard generalizer, the \texttt{class}, cannot
easily be used as we must be able to invalidate cache entries upon
class redefinition.  The issue of \texttt{class} generalizers we can
solve as in \cite{Kiczales.Rodriguez:1990} by using the \texttt{wrapper}
of a class, which is distinct for each distinct (re)definition of
a class; for arbitrary generalizers, however, there is \emph{a priori}
no good way of computing a suitable hash key automatically, so we
allow the metaprogrammer to specify one by defining a method on
\texttt{generalizer-equal-hash-key}, and combining the hash keys for all
required arguments in a list to use as a key in an \texttt{equal}
hash-table.

\subsection{Performance}
\label{sec-3-2}
We have argued that the protocol presented here allows for
expressive control of method dispatch while preserving the
possibility of efficiency.  In this section, we quantify the
efficiency that the memoization protocol described in section
\ref{sec-3-1-2} achieves, by comparing it both to the same protocol
with no memoization, as well as with equivalent dispatch
implementations in the context of methods with regular specializers
(in an implementation similar to that in
\cite{Kiczales.Rodriguez:1990}), and with implementation in
straightforward functions.  We performed our benchmarks on a
quad-core X-series ThinkPad with 8GB of RAM running Debian
GNU/Linux, and took the mean of the 10 central samples of 20 runs,
with the number of iterations per run chosen so as to take
substantially over the clock resolution for the fastest case.
Despite these precautions, we advise against reading too much into
these numbers, which are best used as an order-of-magnitude
estimate.

In the case of the \texttt{cons-specializer}, we benchmark the walker
acting on a small but non-trivial form.  The implementation
strategies in the table below refer to: an implementation in a
single function with a large \texttt{typecase} to dispatch between all the
cases; the natural implementation in terms of a standard generic
function with multiple methods (the method on \texttt{cons} having a
slightly reduced \texttt{typecase} to dispatch on the first element, and
other methods handling \texttt{symbol} and other atoms); and three
separate cases using \texttt{cons-specializer} objects.  As well as
measuring the effect of memoization against the full invocation
protocol, we can also introduce a special case: when only one
argument participates in method selection (all the other required
arguments only being specialized on \texttt{t}), we can avoid the
construction of a list of hash keys and simply use the key
from the single active generalizer directly.

\begin{center}
\begin{tabular}{lrl}
implementation & time (µs/call) & overhead\\
\hline
function & 3.17 & \\
standard-gf/methods & 3.6 & +14\%\\
cons-gf/one-arg-cache & 7.4 & +130\%\\
cons-gf & 15 & +370\%\\
cons-gf/no-cache & 90 & +2700\%\\
\end{tabular}
\end{center}

The benchmarking results from this exercise are promising: in
particular, the introduction of the effective method cache speeds
up the use of generic specializers in this case by a factor of 6,
and the one-argument special case by another factor of 2.  For this
workload, even the one-argument special case only gets to within a
factor of 2-3 of the function and standard generic function
implementations, but the overall picture is that the memoizability
in the protocol does indeed drastically reduce the overhead
compared with the full invocation.

For the \texttt{signum-specializer} case, we choose to benchmark the
computation of 20!, because that is the largest factorial whose
answer fits in SBCL's 63-bit fixnums – in an attempt to measure the
worst case for generic dispatch, where the work done within the
methods is as small as possible without being meaningless, and in
particular does not cause heap allocation or garbage collection to
obscure the picture.

\begin{center}
\begin{tabular}{lrl}
implementation & time (µs/call) & overhead\\
\hline
function & 0.6 & \\
standard-gf/fixnum & 1.2 & +100\%\\
signum-gf/one-arg-cache & 7.5 & +1100\%\\
signum-gf & 23 & +3800\%\\
signum-gf/no-cache & 240 & +41000\%\\
\end{tabular}
\end{center}

The relative picture is similar to the \texttt{cons-specializer} case;
including a cache saves a factor of 10 in this case, and another
factor of 3 for the one-argument cache special case.  The cost of
the genericity of the protocol here is starker; even the
one-argument cache is a factor of 6 slower than the standard
generic-function implementation, and a further factor of 2 away
from the implementation of factorial as a function.  We discuss
ways in which we expect to be able to improve performance in
section \ref{sec-5-1}.

We could allow the metaprogrammer to improve on the one-argument
performance by constructing a specialized cache: for \texttt{signum}
arguments of \texttt{rational} arguments, the logical cache structure is
to index a three-element vector with \texttt{(1+ signum)}.  The current
protocol does not provide a way of eliding the two generic function
calls for the generic cache; we discuss possible approaches in
section \ref{sec-5}.

\subsection{Full protocol}
\label{sec-3-3}
The protocol described in this paper is only part of a complete
protocol for \texttt{specializer} and \texttt{generalizer} metaobjects.  Our
development of this protocol is as yet incomplete; the work
described here augments that in \cite{Newton.Rhodes:2008}, but is
yet relatively untested – and additionally our recent experience of
working with that earlier protocol suggests that there might be
useful additions to the handling of \texttt{specializer} metaobjects,
independent of the \texttt{generalizer} idea presented here.
\section{Related Work}
\label{sec-4}

The work presented here builds on specializer-oriented programming
described in \cite{Newton.Rhodes:2008}.  Approximately
contemporaneously, filtered dispatch \cite{Costanza.etal:2008} was
introduced to address some of the same use cases: filtered dispatch
works by having a custom discriminating function which wraps the
usual one, where the wrapping function augments the set of
applicable methods with applicable methods from other (hidden)
generic functions, one per filter group; this step is not memoized,
and using \texttt{eql} methods to capture behaviours of equivalence classes
means that it is hard to see how it could be.  The methods are then
combined using a custom method combination to mimic the standard
one; in principle implementors of other method combinations could
cater for filtered dispatch, but they would have to explicitly
modify their method combinations.  The Clojure programming language
supports multimethods\footnote{\url{http://clojure.org/multimethods}} with a variant of filtered dispatch as
well as hierarchical and identity-based method selectors.

In context-oriented programming
\cite{Hirschfeld.etal:2008,Vallejos.etal:2010}, context dispatch
occurs by maintaining the context state as an anonymous class with
the superclasses representing all the currently active layers; this
is then passed as a hidden argument to context-aware functions.  The
set of layers is known and under programmer control, as layers must
be defined beforehand.

In some sense, all dispatch schemes are specializations of predicate
dispatch \cite{Ernst.etal:1998}.  The main problem with predicate
dispatch is its expressiveness: with arbitrary predicates able to
control dispatch, it is essentially impossible to perform any
substantial precomputation, or even to automatically determine an
ordering of methods given a set of arguments.  Even Clojure's
restricted dispatch scheme provides an explicit operator for stating
a preference order among methods, where here we provide an operator
to order specializers; in filtered dispatch the programmer
implicitly gives the system an order of precedence, through the
lexical ordering of filter specification in a filtered function
definition.

The Slate programming environment combines prototype-oriented
programming with multiple dispatch \cite{Salzman.Aldrich:2005}; in
that context, the analogue of an argument's class (in Common Lisp)
as a representation of the equivalence class of objects with the
same behaviour is the tuple of roles and delegations: objects with
the same roles and delegations tuple behave the same, much as
objects with the same generalizer have the same behaviour in the
protocol described in this paper.

The idea of generalization is of course not new, and arises in other
contexts.  Perhaps of particular interest is generalization in the
context of partial evaluation; for example, \cite{Ruf:1993}
considers generalization in online partial evaluation, where sets of
possible values are represented by a type system construct
representing an upper bound.  Exploring the relationship between
generalizer metaobjects and approximation in type systems might
yield strategies for automatically computing suitable generalizers
and cache functions for a variety of forms of generalized dispatch.

\section{Conclusions}
\label{sec-5}
In this paper, we have presented a new generalizer metaobject
protocol allowing the metaprogrammer to implement in a
straightforward manner metaobjects to implement custom method
selection, rather than the standard method selection as standardized
in Common Lisp.  This protocol seamlessly interoperates with the
rest of CLOS and Common Lisp in general; the programmer (the user of
the custom specializer metaobjects) may without constraints use
arbitrary method combination, intercede in effective method
combination, or write custom method function implementations.  The
protocol is expressive, in that it handles forms of dispatch not
possible in more restricted dispatch systems, while not suffering
from the indeterminism present in predicate dispatch through the use
of explicit ordering predicates.

The protocol is also reasonably efficient; the metaprogrammer can
indicate that a particular effective method computation can be
memoized, and under those circumstances much of the overhead is
amortized (though there remains a substantial overhead compared with
standard generic-function or regular function calls).  We discuss
how the efficiency could be improved below.
\subsection{Future work}
\label{sec-5-1}
Although the protocol described in this paper allows for a more
efficient implementation, as described in section \ref{sec-3-1-2},
than computing the applicable and effective methods at each generic
function call, the efficiency is still some way away from a
baseline of the standard generic-function, let alone a standard
function.  Most of the invocation protocol is memoized, but there
are still two full standard generic-function calls –
\texttt{generalizer-of-using-class} and \texttt{generalizer-equal-hash-key} – per
argument per call to a generic function with extended specializers,
not to mention a hash table lookup.

For many applications, the additional flexibility afforded by
generalized specializers might be worth the cost in efficiency, but
it would still be worth investigating how much the overhead from
generalized specializers can be reduced; one possible avenue for
investigation is giving greater control over the cacheing strategy
to the metaprogrammer.

As an example, consider the \texttt{signum-specializer}.  The natural
cache structure for a single argument generic function specializing
on \texttt{signum} is probably a four-element vector, where the first
three elements hold the effective methods for \texttt{signum} values of
-1, 0, and 1, and the fourth holds the cached effective methods for
everything else.  This would make the invocation of such functions
very fast for the (presumed) common case where the argument is in
fact a real number.  We hope to develop and show the effectiveness
of an appropriate protocol to allow the metaprogrammer to construct
and exploit such cacheing strategies, and (more speculatively) to
implement the lookup of an effective method function in other ways.

We also aim to demonstrate support within this protocol for some
particular cases of generalized specializers which seem to have
widespread demand (in as much as any language extension can be said
to be in “demand”).  In particular, we have preliminary work
towards supporting efficient dispatch over pattern specializers
such as implemented in the \textsf{Optima} library\footnote{\url{https://github.com/m2ym/optima}}, and over
a prototype object system similar to that in Slate
\cite{Salzman.Aldrich:2005}.  Our current source code for the work
described in this paper can be seen in the git source code
repository at \url{http://christophe.rhodes.io/git/specializable.git},
which will be updated with future developments.

Finally, after further experimentation (and, ideally, non-trivial
use in production) if this protocol stands up to use as we hope, we
aim to produce a standards-quality document so that other
implementors of Common Lisp can, if they choose, independently
reimplement the protocol, and so that users can use the protocol
with confidence that the semantics will not change in a
backwards-incompatible fashion.

\subsection{Acknowledgments}
\label{sec-5-2}
We thank the anonymous reviewers for their helpful suggestions and
comments on the submitted version of this paper.  We also thank Lee
Salzman, Pascal Costanza and Mikel Evins for helpful and
informative discussions, and all the respondents to the first
author's request for imaginative uses for generalized specializers.

\bibliographystyle{plain}
\bibliography{crhodes,specializers}
\end{document}